# Dynamics of quarks and gauge fields in the lowest-energy states in QCD and QED

**Andrew V. Koshelkin**[a] **and Cheuk-Yin Wong**[b]

[a]*National Research Nuclear University - MEPhI, Kashirskoye shosse, 31, 115409 Moscow, Russia*
[b]*Physics Division, Oak Ridge National Laboratory, Oak Ridge, TN 37831, U.S.A.*

E-mail: [a] and.kosh59@gmail.com, [b] wongc@ornl.gov

The dynamics of quarks and gauge fields in the lowest energy states in QCD and QED interactions is studied by compactifying the (3+1)D space-time to the (1+1)D space-time with cylindrical symmetry and by combining Schwinger's longitudinal confinement in (1+1)D with Polyakov's transverse confinement in (2+1)D. Using the action integral, we separate out the transverse and longitudinal degrees of freedom. By solving the derived transverse and longitudinal equations, we study the QCD and QED collective excitations. In addition to the well known QCD low-energy states, we find stable collective QED excitations showing up as massive QED-confined mesons, in support of previous studies. In particular, the masses of the recently observed X17 particle at about 17 MeV and the E38 particle at about 38 MeV are calculated in the developed approach, in good agreement with experimental results.

## 1. Introduction

One of the most exciting problems of modern physics is the origin and nature of the dark matter (DM). There are various DM models, proposing the existence of different hypothetical particles. For example, there are the model assuming the weakly interacting massiive particles (WIMPs) [1, 2], the axions and axion-like particles [3, 4], the dark photons [5–7], the sterile neutrinos [8–10], the fifth force of Nature [11], and the QED-confined composite $q\bar{q}$ mesons and QED neutron [12–14], *etc*.

The DM model based on the QED-confined $q\bar{q}$ mesons is however subject to the question whether a quark and an antiquark can indeed be confined in the QED interaction. It is well-known that in static quark lattice gauge calculations in (3+1)D, quarks are deconfined in compact QED because the static quark compact QED lies in the weak-coupling regime [15]. On the other hand, according to Schwinger, there is a confined regime in (1+1)D QED for dynamical massless quarks [16], and according to Polyakov, there is also a confined regime for quarks in (2+1)D compact QED [17], for all gauge coupling interaction strengths. Furthermore, the proposed phenomenological QED-confined $q\bar{q}$ composite particles in [12, 13] are supported by experimental observations of the X17 [18], the E38 [19], and the anomalous soft photons [20]. Therefore, it is worth examining whether there can be an additional theoretical support for the proposed QED-confined $q\bar{q}$ mesons [12, 13] by combining the longitudinal confinement of Schwinger's massless dynamical quarks in (1+1)D QED with Polyakov's transverse confinement of quarks in compact QED in (2+1)D.

We take Polyakov's transverse confinement configuration in compact QED in (2+1)D as input to construct a "stretch (2+1)D" flux-tube model for the production of a quark and an antiquark in a QED meson in (3+1)D space-time [21]. The Polyakov's transverse confinement of the quark and the antiquark is realized on the transverse $(x^1, x^2)$-plane at $x^3 \sim 0$, at the birth of the $q$-$\bar{q}$. The creation of the charge $q$-$\bar{q}$ pair will be accompanied by the associated creation of their confining gauge fields $\boldsymbol{A}, \boldsymbol{E}$, and $\boldsymbol{B} = \nabla \times A$, which by causality can only be in the neighborhood of the created charges initially with the created $\boldsymbol{E}$ and $\boldsymbol{B}$ fields lie along the longitudinal $x^3$ direction. Subsequent to their birth, the quark and antiquark will execute stretching and contracting "yo-yo" motion along the longitudinal $x^3$ direction, appropriate for the QED meson bound state we are studying. As the quark and antiquark stretch outward in the longitudinal $x^3$ directions, we can construct a longitudinal tube structure of gauge fields in the stretch (2+1)D configuration by duplicating longitudinally the transversely-confined gauge fields that exist on the transverse $(x^1, x^2)$-plane at $x^3 \sim 0$ initially at their birth, for the longitudinal region between the stretching quark and antiquark. The particles, created in such a special (3+1)D space-time can be considered to be as candidates for the QED-confined $q\bar{q}$ composites as DM particles [12, 13]. We examine the model for quarks and gauge fields, keeping in our mind the possibility to compare with experimental results on X17 and E38 bosons [18, 19].

## 2. Reducing (3+1) QCD-QED to (1+1)D QCD-QED with cylindrical symmetry

We start from the $U(1) \bigotimes SU(3)$ Lagrangian for quarks interacting in QCD and QED in (3+1)D space-time in the stretch (2+1)D configuration [22]

$$\mathcal{A}_{4D} = \int d^4 x \, Tr \left\{ \bar{\Psi}(x) \gamma^\mu \Pi_\mu \Psi(x) - \bar{\Psi}(x) m \Psi(x) - \mathcal{L}_A \right\}, \qquad (1)$$

*Dynamics of quarks and gauge fields in lowest states in QCD and QED*

$$\gamma^\mu \Pi_\mu = i\slashed{D} = \gamma^\mu i D_\mu = \gamma^\mu (i\partial_\mu + g_{4D} A_\mu) = \gamma^\mu (p_\mu + g_{4D} A_\mu), \quad (2)$$

$$\mathcal{L}_A = \frac{1}{2\pi^2 R_T^4 g_{4D}^2}[1 - \cos(\pi R_T^2 g_{4D} F_{\mu\nu}(x))], \quad (3)$$

$$F_{\mu\nu} = \partial_\mu A_\nu - \partial_\nu A_\mu - i g_{4D}[A_\mu, A_\nu], \quad F_{\mu\nu} = F^a_{\mu\nu} t_a, \quad A_\mu = A^a_\mu t_a. \quad (4)$$

In these equations we introduce the quark fields $\Psi$, and the QCD and QED gauge fields $A^\mu_{\{QCD,QED\}}$. The QCD gauges fields are $A^\mu_{QCD} = \sum_{a=1}^{8} A^\mu_a t^a$, where $\{t^1, t^2, t^3, ..., t^8\}$ are the SU$_c$(3) generators, $\mu = 0, 1, 2, 3$ are the indices of the space time coordinates $x^\mu$, with the signature $g^{\mu\mu} = (1, -1, -1, -1)$. The QED gauge field is $A^\mu_{QED} = A^\mu_0 t^0$ where $t^0$ is the generator of the color-singlet U(1) subgroup,

$$t^0 = \frac{1}{\sqrt{6}}\begin{pmatrix} 1 & 0 & 0 \\ 0 & 1 & 0 \\ 0 & 0 & 1 \end{pmatrix}. \quad (5)$$

We shall use the summation convention over repeated indices, except when the summation symbols are needed to resolve ambiguities. Here the subscript label of '4D' in $g_{4D}$ and $\mathcal{A}_{4D}$ is to indicate that $g_{4D}$ is the coupling constant in 4D space-time, $\mathcal{A}_{4D}$ is the action integral over the 4D space-time coordinates of $x^0, x^1, x^2$, and $x^3$, and $m$ is the quark mass. The quantity $\sqrt{\pi} R_T$ in Eq. (3) is a transverse length scale which has been chosen to be the square root of the flux tube area.

We reduce (3+1)D QCD-QED to (1+1)D, by taking the fermion fields with cylindrical symmetry in the stretch (2+1)D configuration in the form

$$\Psi_{4D} = \Psi(x) = \begin{pmatrix} G_1(\boldsymbol{r}_\perp) f_+(X) \\ G_2(\boldsymbol{r}_\perp) f_-(X) \\ G_1(\boldsymbol{r}_\perp) f_-(X) \\ -G_2(\boldsymbol{r}_\perp) f_+(X) \end{pmatrix}, \quad \text{where } \boldsymbol{r}_\perp = (x^1, x^2) \text{ and } X = (x^3, x^0). \quad (6)$$

After some bulky manipulations separating the longitudinal and transverse motions of both the quark fields and the gauge fields, we obtain the (1+1)D Lagrangian with cylindrical flux-tube symmetry [22]

$$\begin{aligned}\mathcal{A}_{2D} &= Tr\int dX \left\{ \bar{\psi}(X) \gamma^\mu_{2D} (p_\mu + g_{2D} A_\mu(X)) \psi(X) - m_T \bar{\psi}(X) \psi(X) \right\} \\ &\quad - \int dt\, (\kappa_1 + \kappa_2) |x^3(\bar{q}) - x^3(q)| + Tr \int dX \left\{ -\frac{1}{2} F_{03}(X) F^{03}(X) \right\}, \end{aligned} \quad (7)$$

where $g_{2D}$, $m_T$, $\kappa_1$ and $\kappa_2$ are given in [22]. Note that $g_{2D}$ is a dimensional coupling constant.

The transverse motion is governed by the magnetic field-like equations whose solutions for fermions have the standard form corresponding to the Landau level states for the stretch (2+1)D configuration

$$G_1(\boldsymbol{r}_\perp) = C_1 e^{i\Lambda_1 \phi} e^{-r_\perp^2/2} r_\perp^{|\Lambda_1|} L_n^{(|\Lambda_1|)}(r_\perp^2), \quad (8a)$$

$$G_2(\boldsymbol{r}_\perp) = C_2 e^{i\Lambda_2 \phi} e^{-r_\perp^2/2} r_\perp^{|\Lambda_2|} L_n^{(|\Lambda_2|)}(r_\perp^2). \quad (8b)$$

In this way, both the 2D gauge fields and the coupling constant are governed by the functions

$G_{1,2}(\boldsymbol{r}_\perp)$ so that

$$A^\mu_{4D}(\boldsymbol{r}_\perp, X) = \frac{g_{4D}}{g_{2D}}\left[G_1^*(\boldsymbol{r}_\perp)G_1(\boldsymbol{r}_\perp) + G_2^*(\boldsymbol{r}_\perp)G_2(\boldsymbol{r}_\perp)\right]A^\mu_{2D}(X), \quad \mu = 0, 3. \tag{9}$$

As a consequence, the coupling constant $g_{2D}$ and $g_{4D}$ are related by

$$g^2_{2D} = g^2_{4D}\int d\boldsymbol{r}_\perp \left[G_1(\boldsymbol{r}_\perp)^*G_1(\boldsymbol{r}_\perp) + G_2(\boldsymbol{r}_\perp)^*G_2(\boldsymbol{r}_\perp)\right]^2, \tag{10}$$

yielding a relation between $g_{2D}$, $g_{4D}$, and the flux tube radius $R_T$ [12, 13]. For our case in the lowest energy transverse zero mode with Polyakov's transverse confinement, the transverse mass $m_T$ of the quark and antiquark is unchanged from their rest mass. The Lagrangian (7) generates 2D Dirac equation for longitudinal motion in (1+1)D,

$$\{\gamma^\mu_{2D}(p_\mu + g_{2D}A^{2D}_\mu(X)) - m_T\}\psi_{2D}(X) = 0, \quad \mu = 0, 3, \tag{11}$$

which for massless quarks results in a gauge-invariant relation between the quark current $j^\mu(X)$ and the gauge field $A^\mu(X) \equiv A^\mu_{2D}(X)$,

$$j^\mu(X) = -\frac{g_{2D}}{\pi}\left(A^\mu(X) - \partial^\mu\frac{1}{\partial_\eta\partial^\eta}\partial_\nu A^\nu(X)\right). \tag{12}$$

## 3. Massive bosons

The action integral (7) in (1+1)D also yields the Maxwell equation,

$$\partial_\nu\{\partial^\nu A^\mu(X) - \partial^\mu A^\nu(X)\} = g_{2D}j^\mu(X), \quad \mu, \nu = 0, 3. \tag{13}$$

The above Maxwell equation and the gauge-invariant relation (12) between the current $j^\mu(X)$ and the gauge field $A^\mu(X)$ lead to the Klein-Gordon equation for both the gauge fields and the currents,

$$\partial_\nu\partial^\nu A^\mu(X) = -\frac{g^2_{2D}}{\pi}A^\mu(X), \quad \text{and} \quad \partial_\nu\partial^\nu j^\mu(X) = -\frac{g^2_{2D}}{\pi}j^\mu(X), \tag{14}$$

which coincide with the Klein-Gordon equation for a boson whose square mass is $m^2_{\text{boson}} = g^2_{2D}/\pi$.

With the initial symmetry $U(1)\bigotimes SU(3)$, there are two modes for both the currents, and the QCD and QED gauge fields. Therefore, there are two kind of boson masses in our case, $m_{\text{QED}}$ and $m_{\text{QCD}}$. As it has been said above, we consider quark situations to test our model by experimental results [18, 19]. Characterizing the observable particle state of isospin $I$ for the $\lambda$ interaction, we have [13, 22]

$$m^2_{\lambda I} = \left[\sum_{f=1}^{N_f} D^\lambda_{If}Q^\lambda_f\right]^2 \frac{4\alpha_{\{\text{QCD,QED}\}}}{\pi R^2_T} + m^2_\pi\frac{\alpha_{\{\text{QCD,QED}\}}}{\alpha_{\text{QCD}}}\frac{\sum_f^{N_f} m_f(D^\lambda_{If})^2}{m_{ud}}, \tag{15}$$

where $D^\lambda_{If}$ is the mixing matrix for isospin $I$, flavor $f$, and interaction $\lambda$ with $\lambda = 0$ for QED and

*Dynamics of quarks and gauge fields in lowest states in QCD and QED*

$\lambda = 1$ for QCD. The calculations in [13, 22] give

|  |  | $[I(J^\pi)]$ | Experimental mass (MeV) | Mass formula Eq. (15) (MeV) |
|---|---|---|---|---|
| QCD meson | $\pi^0$ | $[1(0^-)]$ | 134.9768±0.0005 | 134.9‡ |
|  | $\eta$ | $[0(0^-)]$ | 547.862±0.017 | 498.4±39.8 |
|  | $\eta'$ | $[0(0^-)]$ | 957.78±0.06 | 948.2±99.6 |
| QED meson | X17 | $[0(0^-)]$ | 16.94±0.24# | 17.9±1.5 |
|  | E38 | $[1(0^-)]$ | 37.38±0.71⊕ | 36.4±3.8 |

‡ Calibration mass
# A. Krasznahorkay *et al.*, arxiv:2104.10075
⊕ K. Abraamyan *et al.*, EPJ Web Conf 204,08004(2019)

## 4. Conclusion

Previously, in connection with the anomalous soft photons in hadron productions [20], it was proposed that a quark and an antiquark may be confined and bound by the QED interaction as massive bosons (QED mesons) in the region of many tens of MeV [12]. Although the proposed phenomenological QED-confined $q\bar{q}$ composite particles are supported by experimental observations of the X17, the E38, and the anomalous soft photons [12, 13, 18, 19], lattice gauge calculations indicate on the contrary that a static quark and a static antiquark are not confined in compact QED in (3+1)D because they belong to the weak-coupling deconfined regime [15]. However, such deconfined quarks and antiquarks in compact QED in (3+1)D come from theoretical lattice gauge calculations for a static fermion charges and a static antifermion charges as applied to quarks, and the important Schwinger's longitudinal confinement effect for dynamical light quarks in QED [16] has not been included.

To study the Schwinger's confinement effect for light quarks in QED in (3+1)D, we have constructed a "stretch (2+1)D" flux-tube model [21, 22] by starting from the action integral in (3+1)D with both QCD and QED interactions and assuming Polyakov's transverse confinements in (2+1)D. The dynamics in the (3+1)D space-time can now be separated as the coupling between the transverse (2+1)D and the longitudinal (1+1)D degrees of freedom. We solve the transverse Landau level dynamics to obtain the transverse mass and the transverse wave functions. They enter into the calculation of the coupling constant in the longitudinal dynamics of the idealized (1+1)D. We then idealize the flux tube in the (3+1)D space-time as a one-dimensional string in the (1+1)D space-time. Schwinger's longitudinal confining solution for massless charges in QED in (1+1)D can be applied to our problem of the quark-QED system, and by generalization to the problem of the quark-QCD-QED system.

We find that in the stretch (2+1)D flux-tube model which possesses Schwinger's longitudinal confinement and Polyakov's transverse confinement, there can be stable collective excitations of the quark-QCD-QED systems involving either the QCD or the QED interaction, leading to stable QCD mesons and QED mesons whose masses depend on their coupling constants. They correspond to collective dynamics of the quark-QCD-QED medium executing motion in the color-singlet current and the color-octet current respectively. A phenomenological analysis of the lowest-energy states in the flux-tube model yields agreement with the observed QCD and QED spectra, lending support to the proposed hypotheses of QED mesons in [12, 13]. The developed approach can be used for discovering new candidates for the dark matter representatives.

**Acknowledgments**

The research was supported in part by the Division of Nuclear Physics, U.S. Department of Energy, under Contract No. DE-AC05-00OR22725 with UT-Battelle, LLC. The US government retains and the publisher, by accepting the article for publication, acknowledges that the US government retains a nonexclusive, paid-up, irrevocable, worldwide license to publish or reproduce the published form of this manuscript, or allow others to do so, for US government purposes. DOE will provide public access to these results of federally sponsored research in accordance with the DOE Public Access Plan (http://energy.gov/downloads/doe-public-access-plan), Oak Ridge, Tennessee 37831, USA